\documentclass[aps,pra,twocolumn,english,preprint,10pt,nofootinbib]{revtex4}
\usepackage[T1]{fontenc}
\usepackage[latin9]{inputenc}
\usepackage{babel}
\usepackage{nccmath}

\usepackage{graphicx,amsmath,amssymb,color}
\usepackage[normalem]{ulem}
\usepackage{amsfonts}
\usepackage[toc,page]{appendix}
\usepackage{hyperref}
\usepackage{latexsym}
\usepackage{amsfonts}
\usepackage{algpseudocode}
\usepackage{amsthm}
\usepackage{mathrsfs}
\usepackage{natbib}
\usepackage{color,verbatim}
\usepackage{psfrag}
\usepackage[svgnames]{xcolor}

\preprint{IPPP/20/47} 
\preprint{TUM-HEP-1287-20}

\begin{document}

\title{Cosmological bubble friction in local equilibrium}
\author{Shyam Balaji}
\email{shyam.balaji@sydney.edu.au}
\address{School of Physics, The  University of Sydney, Camperdown NSW 2006, Australia}

\author{Michael Spannowsky}
\email{michael.spannowsky@durham.ac.uk}
\address{Institute for Particle Physics Phenomenology, Department of Physics, Durham University, Durham DH1 3LE, U.K.}

\author{Carlos Tamarit}
\email{carlos.tamarit@tum.de}
\address{Physik-Department T70, Technische Universit\"at M\"unchen, James-Franck Stra{\ss}e 1, D-85748 Garching, Germany}

\begin{abstract}
In first-order cosmological phase transitions, the asymptotic velocity of expanding bubbles is of crucial   relevance for predicting observables like the spectrum of  stochastic gravitational waves, or for establishing the viability of  mechanisms explaining fundamental properties of the universe such as the observed baryon asymmetry. In these dynamic phase transitions, it is generally accepted that subluminal bubble expansion requires out-of-equilibrium interactions with the plasma which are captured by friction terms in the equations of motion for the scalar field. This has been disputed in works pointing out subluminal velocities in local equilibrium arising either from  hydrodynamic effects in deflagrations or from the entropy change across the bubble wall in general situations. We argue that both effects are related and can be understood from the conservation of the entropy of the degrees of freedom in local equilibrium, leading to subluminal speeds for both deflagrations and detonations. The friction effect arises from the background field dependence of the entropy density in the plasma, and can be accounted for by simply imposing local conservation of stress-energy and including field dependent thermal contributions to the effective potential.
We illustrate this with explicit calculations of dynamic and static bubbles for a first-order electroweak transition in a Standard Model extension with additional scalar fields. 

 \end{abstract}

\maketitle

\section{Introduction}

The hot plasma in the early universe may have gone through different phase transitions which contributed to forge the properties of the world around us. Classical examples are the phase transition in Quantum Chromodynamics and, if the temperature at early times was large enough, the electroweak phase transition. Though both of the former are of the crossover type in the Standard Model (SM) \cite{Aoki:2006we,Kajantie:1996mn}, first-order phase transitions remain an intriguing possibility which can be realized in SM extensions. Such transitions,  which proceed through the nucleation and subsequent expansion of bubbles of the thermodynamically preferred phase, are particularly interesting due to the enhanced deviations from equilibrium during the transition. The loss of spatial homogeneity and isotropy due to the colliding bubble walls can source a stochastic background of gravitational waves \cite{Witten:1984rs,Hogan:1986qda}  (see Ref.~\cite{Caprini:2018mtu} for a review) amenable to experimental confirmation by future space-borne interferometers like the Big Bang Observer (BBO)~\cite{Crowder:2005nr}, the Deci-hertz Interferometer Gravitational Wave Observatory 
(DECIGO)~\cite{Seto:2001qf} and LISA~\cite{Audley:2017drz}. On the other hand, if the electroweak phase transition were to be of first-order, the former inhomogeneities coupled with novel CP-violating interactions could lead to the generation of the observed baryon asymmetry through the mechanism of electroweak baryogenesis \cite{Kuzmin:1985mm} (for a review, see Ref.~\cite{Garbrecht:2018mrp}).

The predictions of the physical effects of a first-order phase transition, such as the power emitted in gravitational waves  or the generated baryon asymmetry, crucially depend on the velocity reached by the bubbles expanding through the plasma. While gravitational wave emission is enhanced if the velocity becomes nearly luminal, the generation of the baryon asymmetry requires slow bubbles that allow for the diffusion of the particles reflected  in a CP-violating manner by the advancing bubble. This enables the CP excess in front of the bubble wall to be converted into baryon number asymmetry by sphaleron interactions \cite{Heckler:1994uu}.

For these reasons the estimation of bubble velocities has been the subject of intense study, centered on the understanding of the friction effects between the bubbles and the plasma which may slow the advance of the former. Studies based on kinetic theory \cite{Dine:1992wr,Liu:1992tn,Moore:1995ua,Moore:1995si},  fluctuation-dissipation arguments \cite{Khlebnikov:1992bx,Arnold:1993wc} or non-equilibrium quantum field theory \cite{Konstandin:2014zta} suggest a velocity-dependent friction force caused by deviations from equilibrium interactions in the vicinity of the bubble wall. While most analyses are based on evaluating the rate of momentum transfer integrated across the bubble wall, the ensuing friction force is usually incorporated into the local equation of motion of the scalar field.  The kinetic-theory approach or equivalent methods provide first-principle estimates of friction effects by using Boltzmann equations to estimate the out-of-equilibrium effects. Investigations mostly focusing on SM extensions have recently been performed ~\cite{John:2000zq,Bodeker:2009qy,Konstandin:2014zta,Kozaczuk:2015owa,Hoeche:2020rsg}. Many studies consider an effective friction term proportional to a phenomenological friction parameter $\eta$ \cite{Ignatius:1993qn,KurkiSuonio:1996rk,Espinosa:2010hh}, which is sometimes fixed to match the results from the Boltzmann approach \cite{Megevand:2009ut,Megevand:2009gh,Huber:2011aa,Huber:2013kj}.

The general expectation is that there is no friction in local equilibrium \cite{Turok:1992jp}. Furthermore, it has been argued that the friction force saturates at leading order for high-velocities, such that near-luminal bubble propagation, or ``runaway'' behaviour is a generic possibility \cite{Bodeker:2009qy,Bodeker:2017cim}. This was first disputed in \cite{Konstandin:2010dm}, in which it was argued that hydrodynamic effects in deflagrations can lead to a heating of the plasma in front of the bubble wall, which affects the force driving the expansion of the bubble. More recently, Ref.~\cite{Mancha:2020fzw} showed that in local equilibrium the friction force per unit area follows the relationship 
\begin{align}\label{eq:Force}
\frac{|\vec{F}_{\rm friction}|}{A}=(\gamma^2(v_w)-1)\,T\, |\Delta s|,
\end{align}
 where $\gamma(v_w)$ is the Lorentz contraction factor of the asymptotic bubble wall velocity  $v_w$, and $\Delta s$ the  change in entropy density across the bubble. This force keeps growing with the velocity and prevents the bubbles from runaway behaviour. 

The analysis of Ref.~\cite{Mancha:2020fzw} was based on integrating the stress energy momentum tensor across the bubble wall and assuming a constant temperature and fluid velocity throughout. However, this does not exemplify how friction arises in the local dynamical equations for the scalar field and the plasma, or how to consistently compute both the bubble velocity and the associated entropy change. Furthermore, it was not clarified how the results may be related to the hydrodynamic effects investigated in \cite{Konstandin:2010dm}. In particular, while the latter were expected to only take place in deflagrations, the subluminal speeds found in Ref.~\cite{Mancha:2020fzw} are expected regardless of whether the bubbles expand as deflagrations and detonations.

The goal of this paper is to confirm that indeed local equilibrium is compatible with subluminal bubble expansion, clarify the local origin of the friction forces and the relation to the hydrodynamic effect of Ref.~\cite{Konstandin:2010dm}, and provide consistent estimates of bubble velocities. Rather than arising from additional terms in the scalar's equation of motion, the friction-like behaviour in the presence of local equilibrium is caused by the field-dependence of the local entropy and enthalpy density itself, which enters into the hydrodynamic equations of the plasma. As the scalar bubble expands it enforces local entropy and enthalpy changes in the plasma near the bubble wall, and conservation of stress-energy and the total entropy imply  that the bubble must slow down. We will illustrate this effect quantitatively in an extension of the SM with additional scalars. We estimate bubble-wall velocities both from time-dependent solutions with radial symmetry, or by finding planar solutions to the static equations in the wall frame and matching them to consistent hydrodynamic profiles away from the wall. The latter allows to make contact with the treatment of Ref.~\cite{Konstandin:2010dm}, though as a novelty we find profiles corresponding to subluminal detonations, in accordance with the expectations of Ref.~\cite{Mancha:2020fzw}

The paper is organized as follows. In section~\ref{sec:deqs} we review the differential equations for the scalar field plus plasma, arising simply from imposing the conservation of the stress-energy tensor. Next, in section~\ref{sec:model} we introduce the model used to illustrate the friction-like effects. Section~\ref{sec:NN} presents the results for dynamical deflagration solutions with radial symmetry, while finally in sections~\ref{sec:static} and ~\ref{sec:static2} we consider the asymptotic regime of constant velocity expansion and solve for static bubble profiles in the wall frame compatible with consistent deflagration (section~\ref{sec:static}) and detonation (section~\ref{sec:static2}) solutions of the plasma equations away from the bubble. Finally, conclusions are drawn in section~\ref{sec:conclusions}.

\section{\label{sec:deqs}Differential equations for bubble propagation }
We consider a system involving a real scalar field interacting with a thermal plasma. The stress-energy momentum tensor is given by the sum of contributions from both sectors
\begin{align}
 T^{\mu\nu}= T^{\mu\nu}_\phi+T^{\mu\nu}_{p},
\end{align}
where $\phi$ and $p$ denote the scalar field and the plasma respectively. We assume an ordinary scalar with a potential $V(\phi)$ plus a plasma modelled by a perfect fluid, which can be justified as the leading order approximation in an expansion in terms of gradients of the plasma velocity. As such, we have
\begin{align}\begin{aligned}
  T^{\mu\nu}_\phi=&\,\partial^\mu\phi\partial^\nu\phi-\eta^{\mu\nu}\left(\frac{1}{2}\partial_\rho\phi\partial^\rho\phi-V(\phi)\right),\\
   T^{\mu\nu}_p=&\,  (\rho+p) u^\mu u^\mu -\eta^{\mu\nu} p=\omega\, u^\mu u^\mu -\eta^{\mu\nu} p.
\end{aligned}\end{align}
In the above equations, $u^\mu$ with $\mu=0,1,2,3$ represents the fluid's four-velocity, while $p, \rho$ and $\omega=\rho+p$ correspond to the pressure, energy  density and enthalpy of the plasma. We assume the signature $(+,-,-,-)$ for the Minkowski metric and work in natural units with $c=1$. In terms of the plasma velocity vector $v^i$ with $i=1,2,3$, its magnitude $v\equiv\sqrt{\sum_i (v^i)^2}$ and the Lorentz factor $\gamma(v)=1/\sqrt{1-v^2}$, the 4-velocity can be written as $u^\mu=\gamma(v)(1,v^1,v^2,v^3)$. Covariant conservation of the stress-energy momentum tensor in a cosmological background implies $\nabla_\mu T^{\mu\nu}=0$. Under the typical assumption of a phase transition that proceeds much faster than the Universe's expansion, one may neglect the cosmological scale factor and replace covariant derivatives by ordinary ones. Doing so, the terms in  $\nabla_\mu T^{\mu\nu}$ involving $\partial^\nu\phi$ are proportional to the scalar field's equation of motion in the plasma background and must vanish separately. This yields
\begin{align}\label{eq:deqs}\begin{aligned}
 \Box\phi+\frac{\partial}{\partial \phi}(V(\phi)-p)=&\,0,\\
 \partial_\mu(\omega u^\mu u^\nu-\eta^{\mu\nu}p)+\frac{\partial p}{\partial\phi}\partial^\nu\phi=&\,0.
\end{aligned}\end{align}
As initial time boundary conditions for the plasma, a fluid at rest with a temperature given by the nucleation temperature $T_{\rm nuc}$ at which the bubble formation rate overcomes the Hubble expansion should be considered. For the scalar field, a perturbation of the critical bubble that extremizes the three-dimensional integral of the Lagrangian for static fields should be set as an initial condition.

One recognizes the first equation in Eq.~\eqref{eq:deqs} as the equation of motion of the scalar field at finite temperature. Indeed, under the assumption of local thermal equilibrium with temperature $T$,  the pressure is related to the free energy, which itself is related to the thermal corrections $V_T$ to the effective potential $p=- V_T$.  Hence, we may denote $V(\phi)-p=V(\phi,T)$ and recover the standard equation of motion at finite temperature. Equations equivalent to \eqref{eq:deqs} were obtained in Ref.~\cite{Ignatius:1993qn}, where the authors expressed the total pressure as a radiative contribution proportional to $T^4$ and the additional field dependent terms. We make no such distinction here, thus the simpler notation. Furthermore, the authors of Ref.~\cite{Ignatius:1993qn} added a phenomenological friction term without spoiling stress-energy conservation. This corresponds to substituting the r.h.s. of the two equations in~\eqref{eq:deqs} by  $-\eta\, u^\mu \partial_\mu\phi$ and $\eta \,u^\mu \partial_\mu\phi\partial^\nu\phi$, respectively, where $\eta$ is a friction parameter. 

In the second equation of ~\eqref{eq:deqs}, it should be noted that the terms involving field derivatives of the pressure cancel, but the terms proportional to $\partial\omega/\partial\phi$ survive. Under local thermal equilibrium, one can relate  $\omega$ to the entropy density $s=\omega/T$, so that the  terms proportional to $\partial\omega/\partial\phi$ account for local entropy changes across the bubble wall. It is precisely these terms which give rise to friction-like effects and subluminal bubble propagation. In fact, this connection to entropy changes across the bubble wall matches the result ~\eqref{eq:Force} shown in Ref~\cite{Mancha:2020fzw}. The former approach directly assumed a steady state expansion,  planarity and a common temperature on both sides of the bubble. Our treatment goes beyond the former simplifications by incorporating the friction-like effects at the level of the local field and plasma equations. 

We note that standard thermodynamic identities allow the computation of the entropy density in terms of the pressure or equivalently $V_T$, whose one-loop expression for a general model is a standard result of thermal field theory
\begin{align}\label{eq:thermo}
 \omega(\phi,T)= T \,s= &\,T\,\frac{\partial p}{\partial T}=-T\,\frac{\partial  V_T(\phi,T)}{\partial T}.
\end{align}
This considerably simplifies the calculation of backreaction effects under the assumption of local equilibrium, and allows a quick recovery of the lengthier derivations of entropy in e.g. Ref.~\cite{Mancha:2020fzw}. 

It is worth mentioning that the usual friction terms parameterized by $\eta$ lead to a violation of the conservation of the total entropy of the universe, and thus correspond to out-of-equilibrium, irreversible processes. Indeed, adding the friction term to the second equation in~\eqref{eq:deqs}, contracting with $u_\nu$ and using the thermodynamic identities of Eq.~\eqref{eq:thermo} leads to 
\begin{align}
 \partial_\mu(s u^\mu) =\frac{\eta}{T}\,(u^\mu\partial_\mu\phi)^2.
\end{align}
Integrating the former equation over a region of spacetime between times $t_i$ and $t_f$, applying the divergence theorem and assuming a fluid at rest at the boundary gives $S(t=t_f)-S(t=t_i)=\int d^4x \frac{\eta}{T}\,(u^\mu\partial_\mu\phi)^2$, where $S$ is the total entropy in the spatial volume.\footnote{Note that $s$ is the entropy density in the plasma rest frame, and for a general frame one has to account for the Lorentz contraction in the direction of propagation.} In local equilibrium one expects conservation of $S$, and thus it is consistent to take $\eta=0$. Nevertheless, as we will show in the following sections, friction-like behaviour persists. As the expansion is reversible due to the conservation of entropy, the effective force slowing down the bubble is non-dissipative, and we will refer to it as a backreaction as opposed to a friction force. Its effect will be shown in two ways: by solving the dynamical equations~\eqref{eq:deqs}, and by directly looking for solutions of their static limit so  as to constrain the possible wall velocities \cite{Ignatius:1993qn}. Indeed, a large bubble propagating with constant speed has a steady profile up to subleading curvature effects. As such, static solutions to~\eqref{eq:deqs} that capture the field and fluid near the wall can directly be searched for. For a bubble propagating in the $z$ direction with $v^z\equiv \mathfrak{v}$, the static equations can be written as \cite{Ignatius:1993qn}
\begin{align}\label{eq:static_eqs}\begin{aligned}
 &-\phi''(z)+\frac{\partial}{\partial \phi}(V(\phi,T))=0,\\
 &\omega\gamma^2  \mathfrak{ v}^2+\frac{1}{2}\,(\phi'(z))^2-V(\phi,T)=c_1,\quad
 \omega \gamma^2  \mathfrak{ v}=c_2,
\end{aligned}\end{align}
where $c_1$, $c_2$ are constants which can be traded for the temperature $T_+$ and velocity $v_+$  in front of the bubble wall. We assume a bubble propagating towards positive $z$, so that in the wall frame the fluid velocity $v_+$ is negative.  The last two equations in~\eqref{eq:static_eqs} can be used to express the temperature and velocity in terms of the Higgs field and its derivatives, which then leaves a single equation for the scalar field with a non-standard potential $\hat V(\phi,\phi')= V(\phi,T(\phi,\phi'))$ that depends on $\phi'(z)$. We note that the solutions $T(\phi,\phi'), v(\phi,\phi')$ of the last identities in Eq.~\eqref{eq:static_eqs} can be multi-valued, and due to the quadratic dependence on $\mathfrak{ v}$  and quartic dependence on $T$ one can expect two branches of physical solutions with $T>0$, which we will denote with ``high'' and ``low'', giving larger or smaller values of $|\mathfrak{ v}|$, respectively.
Due to the dependence on $\phi'$, the ``energy'' function 
\begin{align}\label{eq:eps}
 {\cal E}\equiv \frac{1}{2}\phi'(z)^2+\hat V_{\rm high,low}(\phi,\phi')
\end{align} 
is only approximately conserved. The boundary conditions are $\phi'(z)=0, z\rightarrow\pm\infty$, and $\phi\rightarrow0, z\rightarrow\infty$. For numerical calculations one may impose analogous boundary conditions at a finite but large $z$. Given $v_+<0$ and $T_+$,  the former boundary conditions  can be satisfied only for a specific choice of the value $\phi_-(v_+,T_+)$  of the field behind the wall, leading to a prediction of the fluid velocity $v_-(v_+,T_+)$ behind the bubble. On physical grounds, one expects the field far away from the bubble setting into a minimum of the finite-temperature effective potential. Then from  Eq.~\eqref{eq:static_eqs} it follows that one should require $\phi''(z)=0, z\rightarrow\pm\infty$, as enforced in Ref.~\cite{Konstandin:2010dm}. This reduces the ambiguity of the solutions to a single parameter, e.g. $T_+$.

The static solutions for the field, velocity and temperature profiles obtained as before have to be matched to time-dependent profiles away from the bubble wall. Far away in front of the wall, one should recover $T=T_{\rm nuc}$, which fixes the ambiguity of the static solution for the wall once it is matched to a hydrodynamic profile. The time-dependence of the latter is expected because, with the scalar field tending to a constant, the lack of dimensionful scales beyond the temperature in the leading contributions to the plasma equations suggests ``self-similar'' solutions depending on  $
\xi\equiv|\vec{x}|/t$ \cite{Gyulassy:1983rq}. Under this assumption, from the second line in Eq.~\eqref{eq:deqs} one can derive the equation
\begin{align}\label{eq:hydro}\begin{aligned}
 \frac{\xi-v}{\omega}\,\partial_\xi \rho-2\,\frac{v}{\xi}-(1-\gamma^2v(\xi-v))\partial_\xi v=0,\\
 \frac{1-v\xi}{\omega}\,\partial_\xi p-\gamma^2(\xi-v)\partial_\xi v=0.
\end{aligned}\end{align}
The possible types of solutions of the above relativistic fluid equations are well known \cite{Gyulassy:1983rq,Espinosa:2010hh}. One expects two types of solutions: deflagrations --in which the bubble expands with a velocity below the speed of sound in the plasma $c_s^2=\partial_T p/\partial_T \rho$, with the fluid heating up and compressing in front of the bubble and at rest behind it-- and detonations --in which the expansion velocity is above $c_s$, the fluid is unperturbed in front of the bubble, but heats up behind it.

For deflagration profiles, since the fluid is expected to be at rest behind the bubble  one can obtain the wall velocity $v_w$ in the fluid frame from the static wall solution as $v_w=-v_-$. The fluid velocity in front of the bubble in the fluid frame is then obtained from a Lorentz boost as $v_{\rm fluid,+}=(v_+-v_-)/(1-v_+ v_-)$. Together with the temperature $T_+$, this gives boundary conditions for the plasma equations  \eqref{eq:hydro} to be solved in front of the bubble, $T_+$ must be fixed so as to get $T=T_{\rm nuc}$ when the velocity drops to zero in front of the bubble.

For detonation profiles, with the fluid unperturbed in front of the bubble one must impose $T_+=T_{\rm nuc}$. The static wall solution then gives unique boundary conditions $T=T_-$, $v_{\rm fluid,-}=(v_--v_+)/(1-v_+ v_-)$ for Eqs.~\eqref{eq:hydro} behind the bubble. 

 { The task of finding physical solutions can be done as follows: first, one chooses a value of $T_+$ ($=T_{\rm nuc}$ for detonations). For deflagrations, keeping $T_+$ fixed one may scan over different values of $v_+$, solving the scalar equation with the pseudopotential $\hat V$, as well as Eqs.~\eqref{eq:hydro} away from the bubble. For the scalar bubble profile one may use a shooting method, imposing $\phi'(z_{\rm min})=0$, and sampling $\phi(z_{\rm min})$ until one has $\phi(z_{\rm max})\rightarrow0,\phi'(z_{\rm max})\rightarrow0$. By construction, as $\phi=0$ is a local minimum, one will have $\phi''(z_{\rm max})\rightarrow0$. One can then sample different values of $v_+$ until one recovers the nucleation temperature at large $z$. This however does not guarantee that the physical condition $\phi''(z)\rightarrow0$ is attained for small $z$; to achieve this one may repeat the above calculations for different values of $T_+$ until $\phi''(z_{\rm min})$ is minimized. For detonations, with $T_+=T_{\rm nuc}$ one can directly scan for values of $v_+$ until  $\phi''(z_{\rm min})$ is minimized. The approximate conservation of ${\cal E}$ in Eq.~\eqref{eq:eps} can be used to simplify the   search for physical solutions, since if ${\cal E}$ were exactly conserved, the scalar profile of the physical solution would interpolate between two exactly degenerate minima. Hence the physical solutions have values of $T_+,v_+$ for which one gets a pseudopotential $\hat V$ with near degenerate minima.}

In a planar approximation the calculation gets simplified because there is no need to solve \eqref{eq:hydro}. In the planar regime the $1/\xi$ term in Eqs.~\eqref{eq:hydro} can be dropped and one gets solutions with constant velocity and pressure, which simplifies the treatment. However, satisfying the boundary conditions of fluid at rest far from the wall implies the appearance of discontinuity fronts across which the velocity drops to zero: a shock front in front of the bubble in the case of deflagrations, and a similar discontinuity behind the bubble for detonations. One can relate quantities across the front by imposing continuity of the stress-energy tensor. In the case of deflagrations,  equating the fluid velocity between wall and shock front deduced from the solutions of~\eqref{eq:static_eqs} and from the shock constraints leads to the condition
\begin{align}\label{eq:shock}
  \hskip-.1cmv_{\rm fluid,+}=\frac{v_+\!-\!v_-}{1\!-\!v_+v_-}=\frac{\sqrt{3} \left(T_+^4\!-\!T_{\text{nuc}}^4\right)}{\sqrt{\left(T_{\text{nuc}}^4\!+\!3 T_+^4\right) \left(3 T_{\text{nuc}}^4\!+\!T_+^4\right)}}.
\end{align}
{For a given $T_+$, the above can be used to fix the free parameter $v_+$}.
In the case of detonations, there is no additional constraint as one has $T_+=T_{\rm nuc}$, but within the planar approximation the temperature $T_{\rm in}$ inside the bubble beyond the detonation front can be obtained  from the following equations,
\begin{align}\label{eq:det}
 v_{\rm fluid,-}=\frac{v_-\!-\!v_+}{1\!-\!v_+v_-}=\frac{\sqrt{3} \left(T_{\rm in}^4\!-\!T_{-}^4\right)}{\sqrt{\left(T_{\rm in}^4+3 T_-^4\right) \left(3 T_{-}^4+T_{\rm in}^4\right)}}.
\end{align}

%

To make contact with the results of Ref.~\cite{Mancha:2020fzw}, let us point out that the friction force ~\eqref{eq:Force} can be derived directly from the second identity in Eq.~\eqref{eq:static_eqs} evaluated at both sides of the wall (where $\phi'(z)=0$), once one identifies the backreaction pressure ${|\vec{F}_{\rm back}|}/{A}$ with $|\Delta V(\phi,T)|$, and under the approximation of a constant temperature and fluid velocity, the latter identified with $-v_w$. In reality, the situation is more complicated as the temperature and velocity change across the bubble, a more complete result is
\begin{align}\label{eq:FrictionExact}
 \frac{|\vec{F}_{\rm back}|}{A}= |\Delta\{\gamma^2 v^2\omega\}|=  |\Delta\{(\gamma^2-1)T s\}|.
\end{align}
The planar approximation can be used to gain an intuitive understanding of the reasons behind the subluminal propagation speed. In either deflagration or detonation solutions, the interior of the bubble has lower entropy density than the fluid before the transition. This simply follows from the fact that the phase transition makes some particles massive, while the entropy in the plasma is always dominated by the contribution from the relativistic degrees of freedom.  Recall that in thermal plasma, one can write
\begin{align}\label{eq:s}
 s=\frac{2\pi^2}{45}\,g_{\star s} T^3,
\end{align}
where $g_{\star s}$ denotes the number of effective relativistic degrees of freedom.  Inside the bubble $g_{\star s}$ drops, and with it $s$.
For the degrees of freedom in local equilibrium, the total entropy has to be conserved. With the entropy decrease due to the presence and expansion of the bubble, there has to be a compensating entropy increase. Given Eq.~\eqref{eq:s}, this can be achieved if parts of the fluid heat up. This is precisely what happens in detonations and deflagrations, in which the fluid heats up behind and in front of the bubble respectively. In the planar approximation, one simply expects a detonation/deflagration shell with constant increased temperature $T_{\rm shell}$ --corresponding in the notation above to $T_-$/$T_+$ for detonations/deflagrations-- and with an additional shell front propagating with constant velocity $v_{\rm front}$ behind/ahead of the bubble wall.
The conservation of the total entropy within this approximation then gives
\begin{align}\label{eq:vwall_vfront}
 v_{w}=v_{\rm front}\left(\frac{|\Delta \gamma s|_{\rm front}}{|\Delta \gamma s|_{\rm wall}}\right)^{1/3},
\end{align}
where we assumed fluid shells with radial symmetry and radii $R_w=v_wt, R_{\rm front}=v_{\rm front}t$.
 Using the stress-energy conservation relations across the front, one can relate $v_{\rm front}$ to the temperatures at each side of the front,
\begin{align}
 v_{\rm front}=\left\{\begin{array}{cc}
                       \frac{1}{\sqrt{3}}\left(\frac{3T_-^4+T_{\rm in}^4}{3T_{\rm in}^4+T_{-}^4}\right)^{1/2}\quad \text{detonations}\\
                         \frac{1}{\sqrt{3}}\left(\frac{3T_+^4+T_{\rm nuc}^4}{3T_{\rm nuc}^4+T_{+}^4}\right)^{1/2}\quad \text{deflagrations.}
                      \end{array}
\right.
\end{align}
One can also express the entropy increase across the front in terms of the same temperatures using  Eqs.~\eqref{eq:shock}, \eqref{eq:det} and \eqref{eq:s}. Subliminal speeds are generally expected for moderate heating in the compression shell. 

Above, we related the subluminal propagation speeds to a heating effect associated with the conservation of the entropy of the degrees of freedom in local equilibrium. A heating effect was already connected to subliminal speeds in local equilibrium  in the case of deflagrations in Ref.~\cite{Konstandin:2010dm}, though with different argumentation. It was noted that such a heating in front of the bubble wall could lead to a zero driving force, incorporating the effects of pressure and the zero $T$ potential difference, for the bubble expansion. In view of the arguments provided in Ref.~\cite{Mancha:2020fzw} (which, as seen above, follows from the static equations \eqref{eq:static_eqs}, which were also solved in Ref.~\cite{Konstandin:2010dm}), one does not expect an exactly zero driving force, but a compensation with a backreaction force due to the entropy changes across the bubble. Yet the heating effect first noted in  Ref.~\cite{Konstandin:2010dm} is definitely connected with subluminal propagation speeds, and can be understood from entropy conservation and extended to detonations.

\section{\label{sec:model}Example model}
To illustrate the friction effects, we consider an extension of the SM by an $N$-dimensional multiplet $\chi$ of complex scalar singlets with  $U(N)$-preserving couplings, including interactions with the Higgs $\Phi$: 
\begin{eqnarray}
{\cal L} \supset&& -m^2_{H}\Phi^\dagger\Phi-\frac{\lambda}{2}(\Phi^\dagger\Phi)^2-m^2_{\chi}\chi^\dagger\chi \nonumber\\
&&-\frac{\lambda_\chi}{2} (\chi^\dagger\chi)^2-\lambda_{H\chi}\Phi^\dagger\Phi \chi^\dagger\chi \; .
\end{eqnarray}
Now, all that is required for writing down the equations is $p=- V_T$. For simplicity of the numerical implementation we use a high-temperature expansion up to terms of order $T$, which still captures the nontrivial field dependence
\begin{align}\label{eq:pressure}
 \begin{aligned}
 &p(h,T)=\,\frac{\pi ^2 T^4}{90}  (g_{*,\rm SM}+2N)-T^2\left(h^2 \left(\frac{y_b^2}{8}+\frac{3 g_1^2}{160}\right.\right.\\
 &\left.\left.+\frac{3 g_2^2}{32}+\frac{\lambda }{8}+\frac{N \lambda _{H\chi}}{24}+\frac{y_t^2}{8}\right)+\frac{m^2_H}{6}+\frac{N m^2_\chi}{12}\right)\\
 &-\frac{T}{12\pi}\left(-\frac{3}{4} \left(g_2 h\right)^3-\frac{3 h^3}{8} \left(\frac{3 g_1^2}{5}+g_2^2\right)^{3/2}\right.\\
 &-3 \left(\frac{h^2 \lambda }{2}+m^2_H\right)^{3/2}-\left(\frac{3 h^2 \lambda }{2}+m^2_H\right)^{3/2}\\
 &\left.-2 N \left(\frac{h^2 \lambda _{H\chi}}{2}+m^2_\chi\right)^{3/2}\right).
  \end{aligned}
\end{align}
In the above equation, we have assumed a background for the neutral component of the Higgs $h$. $g_{\star,\rm SM}\sim 106.75$ denotes the number of effective relativistic degrees of freedom in the SM plasma, while $g_1$ and $g_2$ are the hypercharge and weak gauge couplings in the normalization compatible with Grand Unification, and  $y_t,y_b$  are the bottom and top quark Yukawa couplings respectively. For the couplings and parameters beyond those of the SM we use $N=4$ or $N=2$, $m^2_S/m^2_W=0.0625$, $\lambda_\chi=0.085$, $\lambda_{H\chi}=0.85$. This gives a first-order electroweak phase transition with critical and nucleation temperatures around $T_c=115.952$ GeV, $T_{\rm nuc}=115.297$ GeV for $N=4$ and $T_c=126.376$ GeV, $T_{\rm nuc}=126.229$ GeV for $N=2$. The nucleation rate can be estimated by minimizing the three-dimensional integral $S_3[h,T]$ of the finite-temperature action evaluated at static configurations $h(r)$ with radial symmetry, we use the standard criterion for nucleation $S_3[h_{\rm nuc}(r),T_{\rm nuc}]/T_{\rm nuc}\sim 140$, where $h_{\rm nuc}(r)$ is the critical field configuration or bubble.  

\section{\label{sec:NN}Solving for time-dependent solutions with a neural network}

In this section we focus on solving the time-dependent equations~\eqref{eq:deqs} for the above parameter choices, with $N=4$. We assume radial symmetry, with the velocity field having a radial component $v^r\equiv v$, and with $v,h,T$  being functions of $r,t$.
As initial conditions we use $T(r,t=0)=T_{\rm nuc}, v(r,t=0)=0$, while for the Higgs we use the critical bubble perturbed with a nonrelativistic boost (as otherwise the bubble would remain static): $h(r,t=0)=h_{\rm nuc}(r),\, \partial_t h(r,t)|_{t=0}=-\delta h'_{\rm nuc}(r)$, with $\delta=0.2$. 

\subsection{\label{subsec:setup} Setup}
In order to find time-dependent solutions to Eqs.~\eqref{eq:deqs} we follow the technique pioneered in Ref. \cite{Piscopo:2019txs} and implement an artificial neural network (NN). The method relies on recasting the partial differential equations (PDEs) as an optimization procedure --for which NN are uniquely suited-- of the form  $\hat{\mathcal{L}}= 0$, where $\hat{\mathcal{L}}$ is a positive loss function to be minimized by the NN. The network is constructed by considering an initial layer of 2 inputs ${\xi}_n=(r,t)$ that are to be mapped to a final layer with 3 outputs $N_m$ which are to be approximations of the solutions $\varphi_m=(v,h,T)$ to the differential equations. The inputs are mapped to successive hidden layers of $k$ elements, from the combined action of linear transformations between each layer and the action of a real activation functions, a final linear mapping gives the final outputs. For example, for one hidden layer one has
\begin{align}\label{eq:NN}
N_m(\vec{\xi},\{w,b\})=\sum_{k,n} w^f_{mk} g(w^h_{kn}\xi_n +b_k^n)+b_m^f,
\end{align} 
where $g$ is the activation function, $\omega^h_{mk},\omega^f_{mk}$ are known as ``weights'', and $b^h, b^f$  are the ``biases''. A set of weights and biases which minimize the loss function associated with the system of differential equations are searched for. Writing the latter in the form
\begin{align}
\label{eq:genericPDE}
{\mathcal{F}}_m(\vec{\xi},\varphi_n(\vec{\xi}), \partial^p_q \phi_n(\vec{\xi}))=0, 
\end{align} 
with $m,n\in\{1,2,3\}$, $p,q\in\{1,2\}$, and assuming boundary conditions (BCs) for boundary points $\vec{\xi}_b$ of the form
\begin{align}
 {\mathcal{B}}_a(\vec{\xi}_b,\varphi_n(\vec{\xi}_b), \partial^p_q \phi_n(\vec{\xi}_b))=0,
\end{align}
the loss function is constructed from considering a discrete set of ``training points'' $\vec{\xi}_i$ including boundary points $\vec{\xi}_{b,j}$, and evaluating  ${\cal F}_m$ and ${\cal B}_a$ on them
\begin{align}
\label{eq:lossfunction}
\hat{\mathcal{L}}(\{w,b\})=&\sum_{i,m} c_m {\mathcal{F}}_m(\vec{\xi}_i,{N}_n(\vec{\xi}_i), \partial^j_k {N}_n(x_i))^2  \nonumber \\
+ \sum_{j,a}& d_a{\mathcal{B}}_a(\vec{\xi}_{b,j},N_n(\vec{\xi}_{b,j}), \partial^j_k N_n(\vec{\xi}_{b,j}))^2.
\end{align} 
Above, the derivatives of the network outputs can be obtained analytically from ~\eqref{eq:NN}. The coefficients $c_m$ and $d_a$ represent relative weightings for each PDE and BC, required to ensure that all PDEs and BCs contribute comparably to the loss function. We implement the NN with 13 hidden layers with $10$ nodes each, with $\tanh$ activation functions. We choose the training examples from an evenly spaced $80 \times 80$ grid. We use the \texttt{pytorch} package along with the \texttt{Adam} optimizer for the NN gradient descent. To avoid getting trapped in sub-optimal local minima of the smooth loss function, we take care to reduce the learning rate through cosine annealing with warm restarts.  For fast convergence of our solution, we first pretrain the NN with a template solution implemented as a boundary condition for low $t$. This is obtained using \texttt{Wolfram Mathematica}'s  PDE solver, which is only able to provide reliable solutions for a small time interval. After the NN is in the correct vicinity of solution, we remove the pretrained template from the loss function and train according to ~\eqref{eq:lossfunction}.  This allows reliable solutions for time intervals that cannot be reached with the \texttt{Mathematica} solver.

\subsection{Dynamic transition results}

\begin{figure}
   \centering
     \hskip.5cm\includegraphics[width=0.32\textwidth]{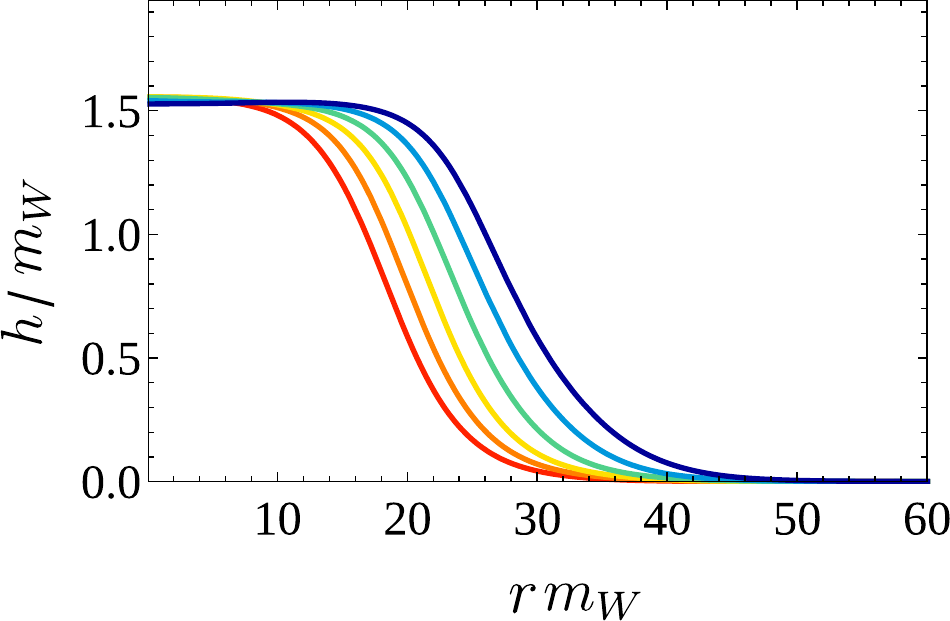}
     \vskip.1cm
     \includegraphics[width=0.35\textwidth]{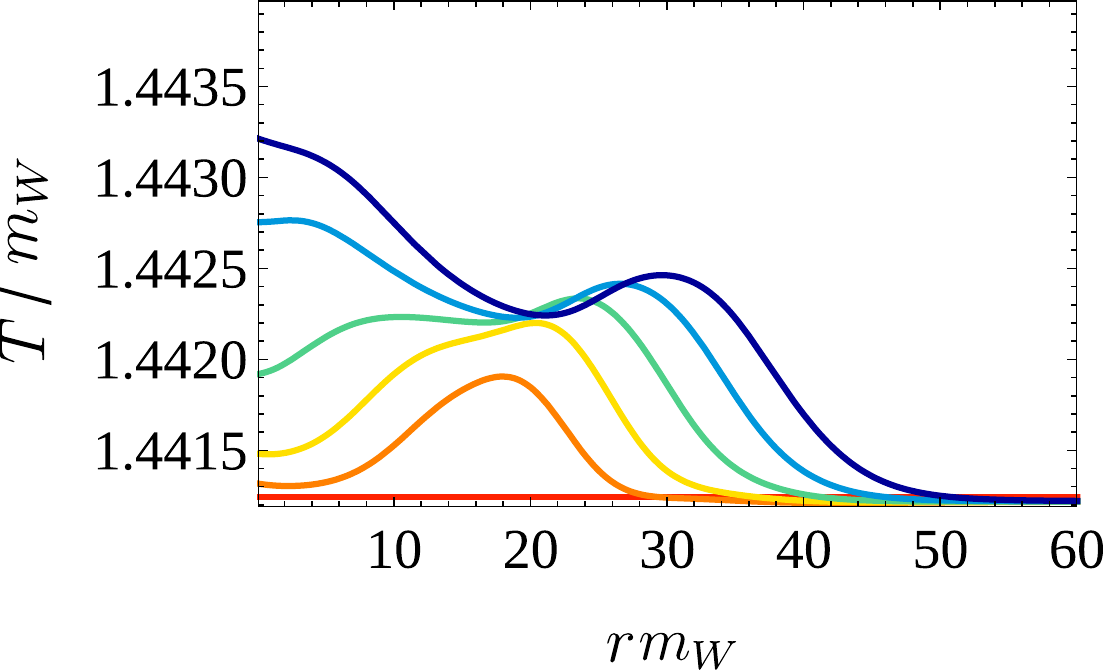}
     \vskip.1cm
   \ \,\hskip-.4cm  \includegraphics[width=0.37\textwidth]{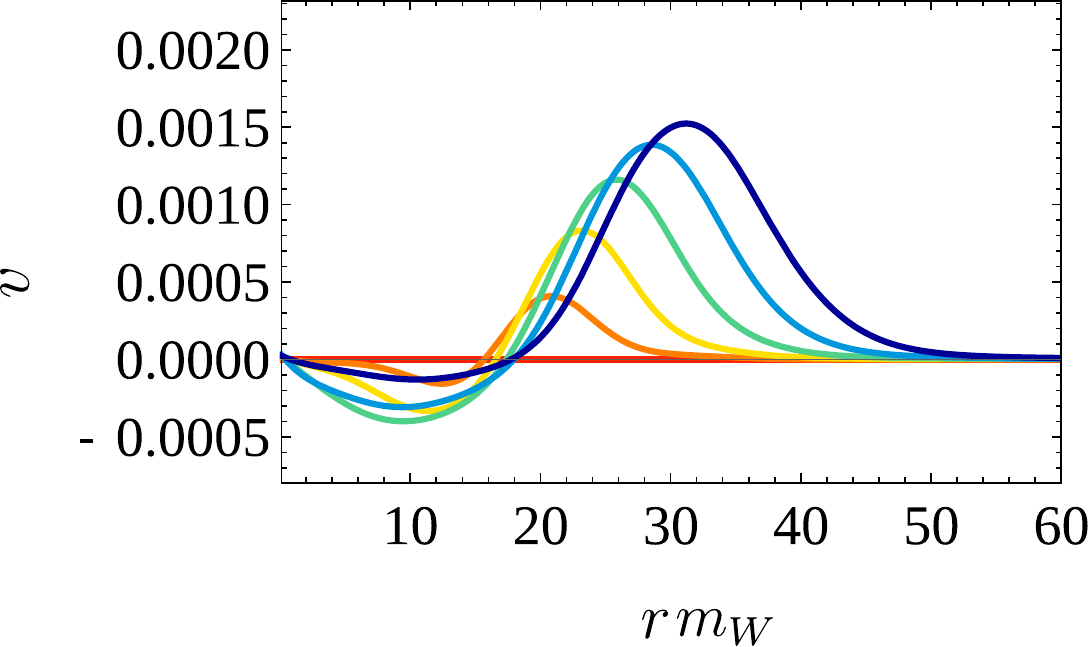}
\caption{Dynamical evolution of $h,T$ and $v$ in dimensionless units. The curves from left to right, red to blue, correspond to  time steps from $t m_W=[0,50]$ with $m_W \Delta t= 10$.}
\label{fig:PhaseTransition}
\end{figure}

\begin{figure}
   \centering
    \hskip0cm \includegraphics[width=0.33\textwidth]{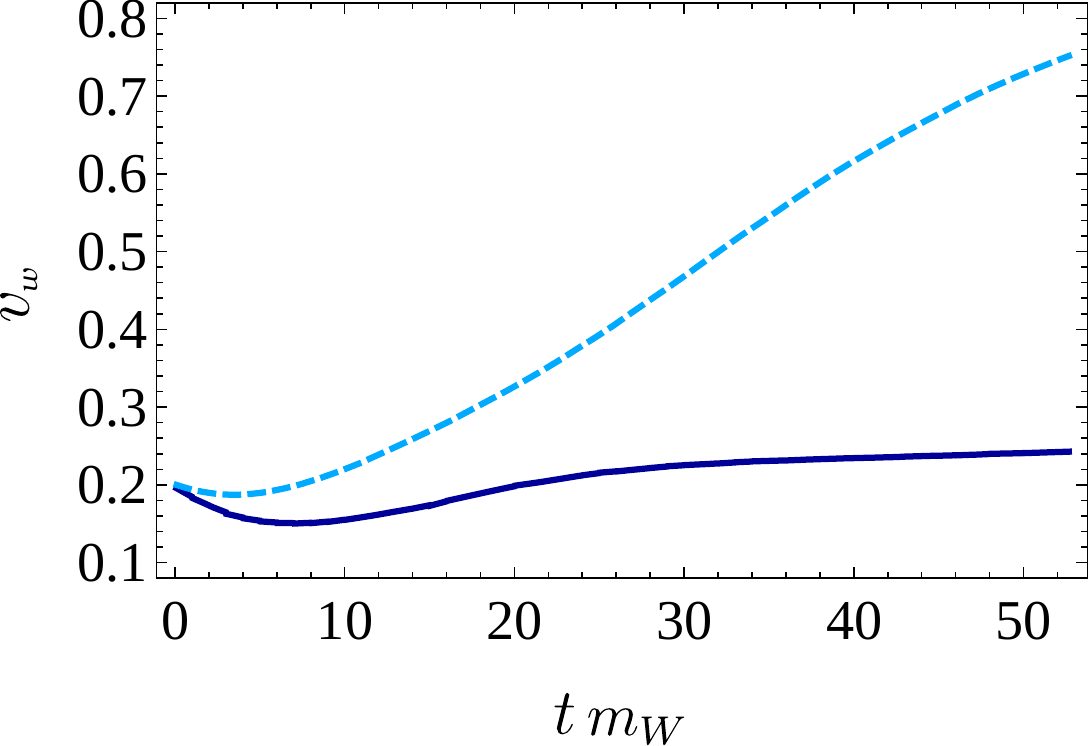}
\caption{Bubble velocity versus time in dimensionless units, including the effect of the field dependence of the enthaply (solid line) or without it (dashed line).}
\label{fig:VelocityEvolution}
\end{figure}

From the previous NN setup we were able to obtain solutions in which the individual loss functions ${\cal F}_m$ in dimensionless units (obtained by rescaling quantities with appropriate powers of the W mass $m_W\approx 80$ GeV) take values  $\lesssim 5 \times 10^{-3}$.
We show the resulting dynamical profiles of $h,T,v$ in Fig.~\ref{fig:PhaseTransition}  as a function of $r$ in  dimensionless units for 5 equally spaced timestamps between $t=0$ and $t=50/m_W$. We note that the \texttt{Mathematica} solver was only able to compute accurate solutions for $t\lesssim15/m_W$. The scalar profile settles to a slow expansion, while the velocity and temperature profiles show the formation of a faster propagating front, in accordance with the expectations of a deflagration solution with self-similar fluid behaviour. Confirming the latter would require extending the solutions to even later times, a more efficient means is to directly look for static wall solutions with consistent hydrodynamic profiles as we show in the next section. By following points with constant $h(r,t)=0.5$ we can estimate the bubble's position and velocity, the latter is plotted with a solid line in Fig.~\ref{fig:VelocityEvolution}, which shows that the bubble's velocity settles to $\lesssim 0.25$. This is in contrast to the result,  illustrated with a dashed line, when the terms proportional to $\partial\omega/\partial\phi$ are omitted in Eq.~\eqref{eq:deqs}. In this case the bubble velocity quickly approaches the speed of light. This confirms our observation that the field-dependence of the enthalpy is responsible for the friction-like behaviour.

\section{\label{sec:static} Static planar bubble profiles and consistent deflagrations}

\begin{figure}
   \centering
     \includegraphics[width=0.34\textwidth]{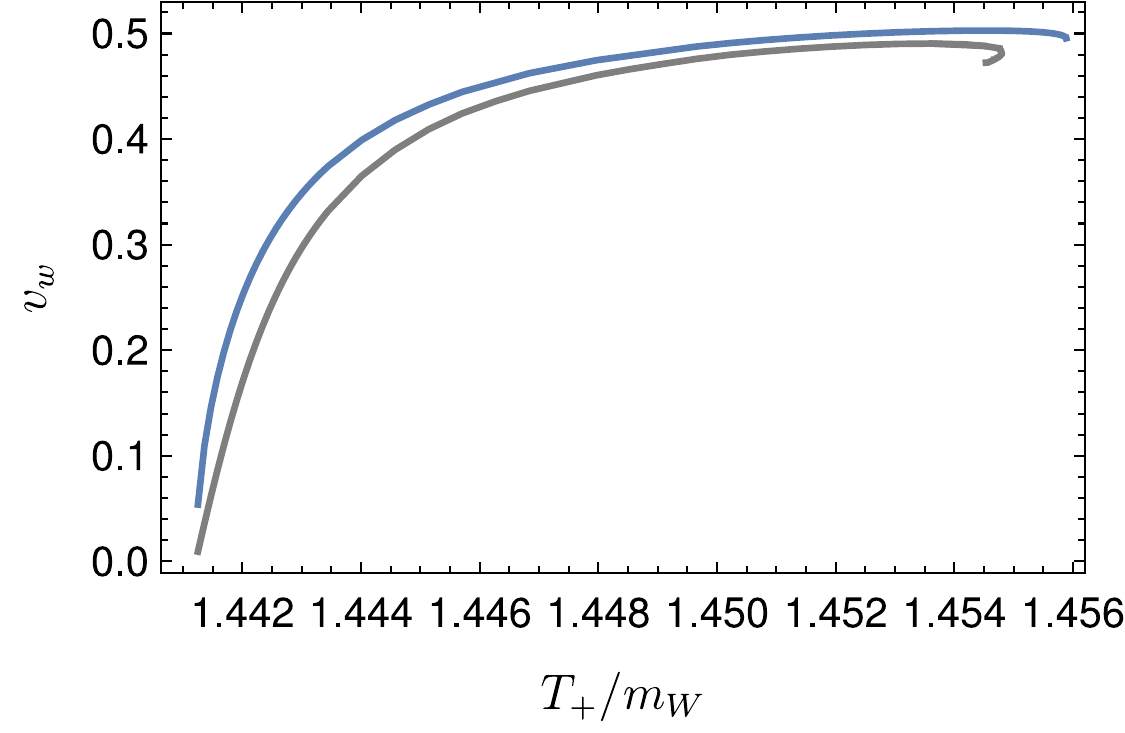}\\
     \vskip.1cm
    \, \hskip-.5cm\includegraphics[width=0.33\textwidth]{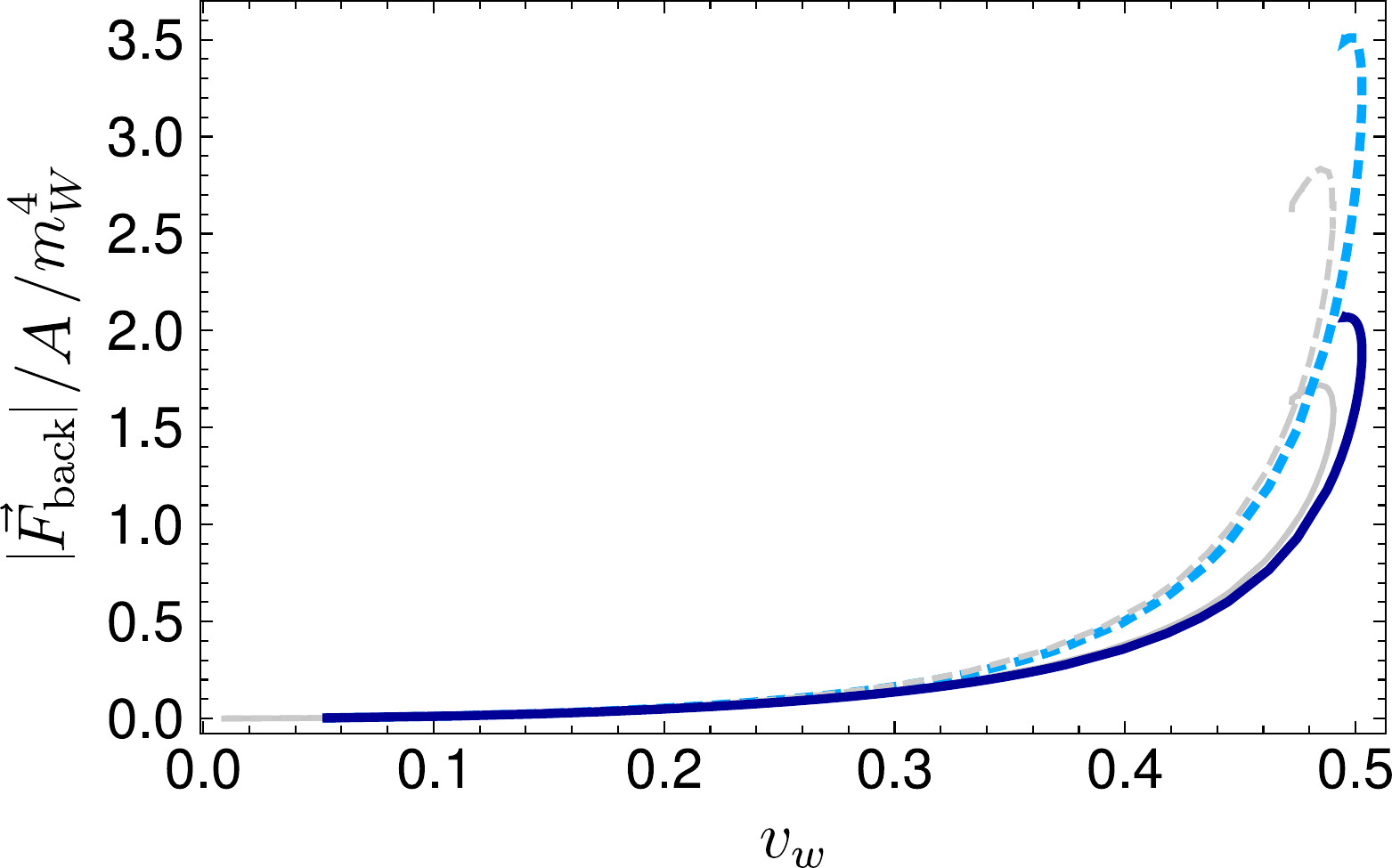}
\caption{\label{Fig:static_results}Upper plot: Bubble velocity as a function of $T_+$ for the static solutions compatible with  consistent deflagrations, and without imposing $\phi''(z)\rightarrow0$ far away from the bubble. The  blue curve gives the results when solving the hydrodynamic equations \eqref{eq:hydro} away from the bubble, while the grey line gives the results in the planar approximation. Lower plot: Backreaction force as a function of the bubble wall velocity (solid lines) compared to its approximation in Eq.~\eqref{eq:Force} (dashed lines). The curves in blue/grey correspond to the hydrodynamic equations with radial/planar symmetry. In both plots, the physical solution with $\phi''(z)\rightarrow0$ at large $|z|$ corresponds to the ending points of the blue curves, or the turning points of the grey curves. }
\label{fig:static_results}
\end{figure}

In this section we report the results of searching for deflagration solutions with the same parameters as in the previous section, assuming a static solution near the bubble wall that solves Eq.~\eqref{eq:static_eqs}, and either implementing the hydrodynamic constraints of Eqs.~\eqref{eq:shock}, \eqref{eq:det} applying in the planar regime, or matching with solutions to the radial hydrodynamic equations \eqref{eq:hydro}. Without imposing the boundary condition $\phi''(z)\rightarrow0$, we have found a one-parameter branch of solutions satisfying all constraints. These family of configurations corresponds to the ``low'' branch of solutions for the temperature profiles $T(h,h')$, and  when solving Eqs.~\eqref{eq:hydro} we find acceptable configurations for $T_+\leq T_+^{
\rm max}=116.471$ GeV. The upper value of $T_+$ corresponds to the unique solution satisfying the physical constraint $\phi''(z)\rightarrow0$ at large $|z|$, and having a wall velocity $v_w=0.496$. We note that with our method it is challenging to exactly recover $\phi''(z_{\rm min})=0$ because we use a finite interval of $z$, and moreover we find an exponential sensitivity of $\phi''(z_{\rm min})$ to the value of $T_+$ near $T_+^{
\rm max}$, with $\phi''(z_{\rm min})$ approaching zero with a slope that seems to grow towards infinity.  The former results are compatible with the dynamical results of the previous section, in which temperatures remained below  the above value of $T_+^{\rm max}$ (see Fig.~\ref{fig:PhaseTransition}) and the wall velocity approached $0.25$. The lower velocity in the dynamical simulation can be due to the effect of considering a radial expansion, as opposed to the planar approximation used for finding the static wall profile. It could also be that the planar wall velocity is only reached at much later times than the ones covered by the dynamical simulation of the previous section, note that the slope of the wall velocity in Fig.~\ref{fig:VelocityEvolution}, though very small at later times, seems to be nonzero. The wall velocity $v_w$, the exact backreaction force of Eq.~\eqref{eq:FrictionExact} and the approximation of Eq.~\eqref{eq:Force} found in Ref.~\cite{Mancha:2020fzw} are illustrated in Fig.~\ref{fig:static_results}, which also shows the results when, instead of solving Eqs.~\eqref{eq:hydro}, one imposes the planar constraints of Eq.~\eqref{eq:shock}. We find qualitative agreement with~Eq.~\eqref{eq:Force} up to deviations below 70\%, which are due to the changes of $T$ and $v$ across the bubble.

\begin{figure}
   \centering
    \hskip0cm \includegraphics[width=0.38\textwidth]{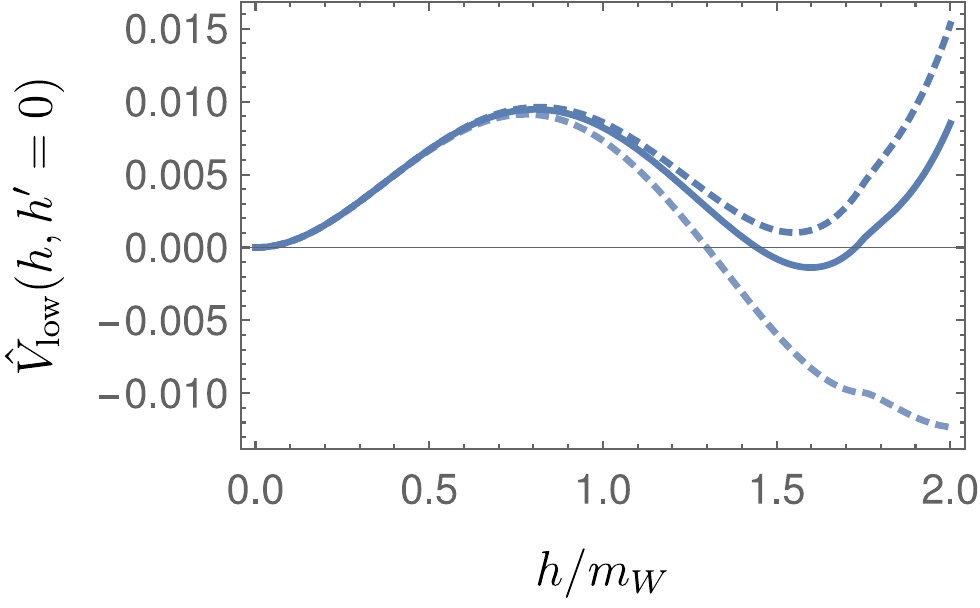}
\caption{Pseudopotential $\hat V_{\rm low}(h,h'=0)$ for $T_+=116.471$ GeV, with $v_+$ taking the values (from top to bottom): -0.47, -0.4809, -0.50. The central choice of $v_+$ gives a hydrodynamic profile in which $T=T_{\rm nuc}$ when the fluid velocity drops to zero, and with a minimal value of $|\phi''(z_{\rm min}=-25/m_W)|$ in our numerical scans.}
\label{fig:Vhatlow}
\end{figure}

\begin{figure}
   \centering
    \hskip0.4cm \includegraphics[width=0.3\textwidth]{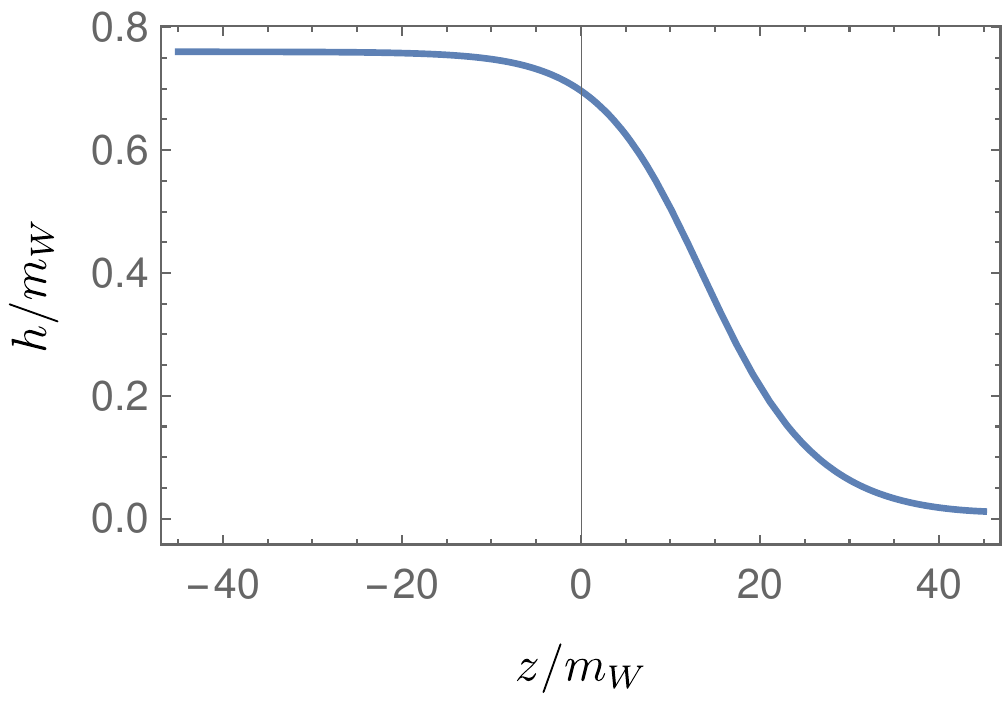}
     \includegraphics[width=0.33\textwidth]{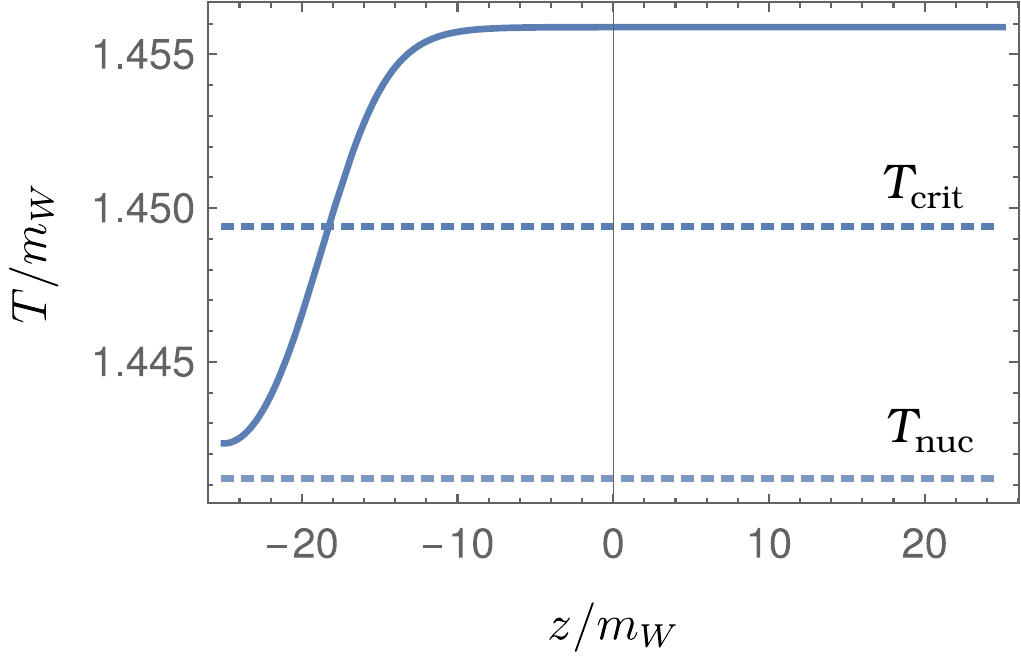}
      \includegraphics[width=0.33\textwidth]{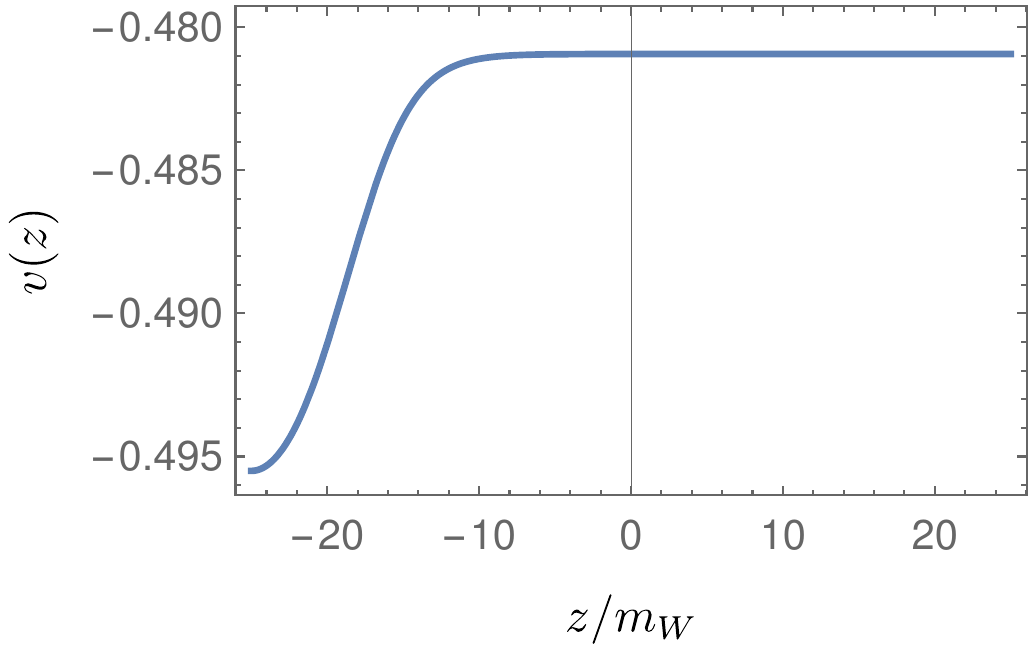}
\caption{Deflagration profiles of the Higgs, temperature and velocity across the bubble wall for $T_+=T_{+}^{\rm max}$.}
\label{Fig:def_prof}
\end{figure}

\begin{figure}
   \centering
     \includegraphics[width=0.33\textwidth]{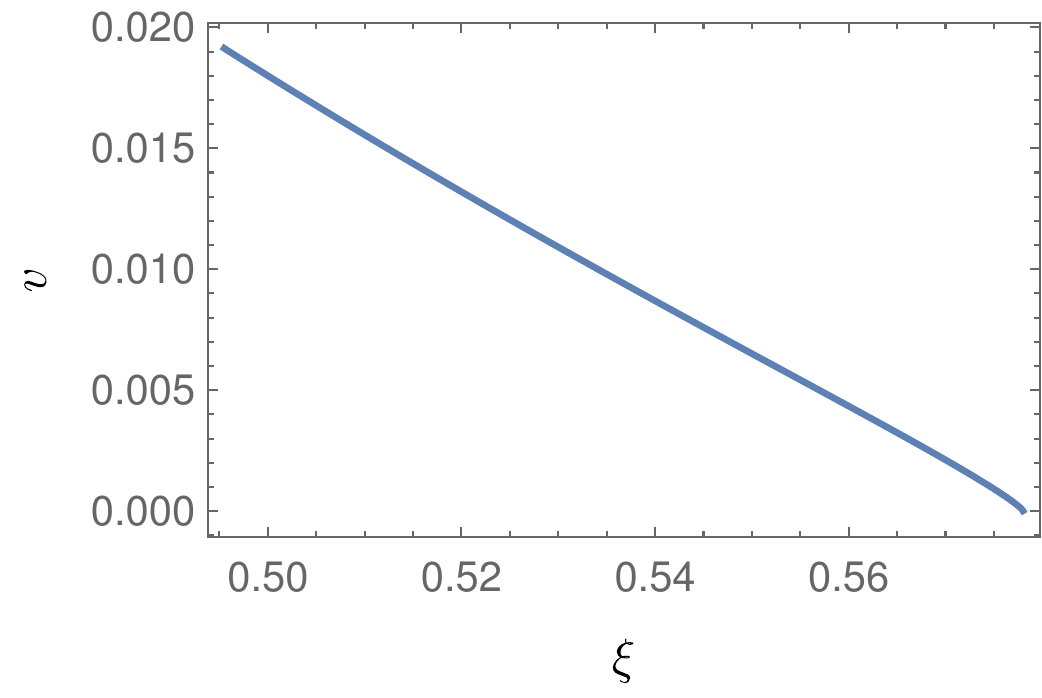}
      \includegraphics[width=0.33\textwidth]{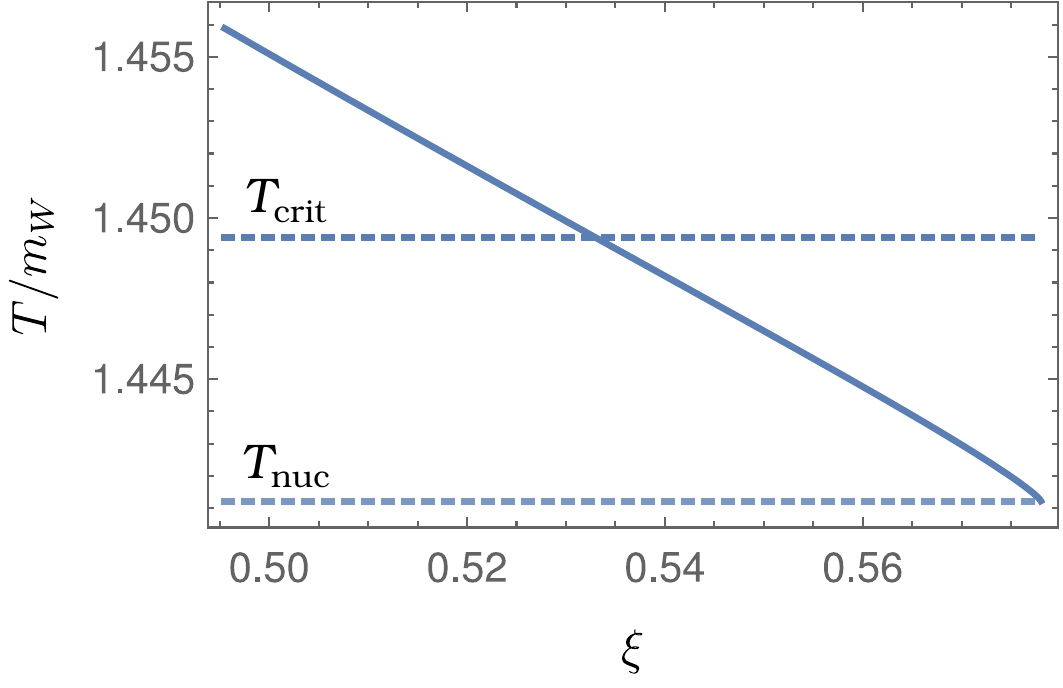}
\caption{Hydrodynamic profiles for the fluid temperature and velocity in front of the bubble wall corresponding to the bubble profiles in Fig.~\ref{Fig:def_prof}.}
\label{Fig:def_prof_xi}
\end{figure}

In Fig.~\ref{fig:Vhatlow} we illustrate the pseudopotential $\hat{V}_{\rm low}(h,h')$ evaluated at constant configurations with $h'=0$, for three different values of $v_+$ and the value of $T_+=116.471$  GeV giving the smallest $|\phi''(z_{\rm min})|$ for $z_{\rm min}=-25/m_W$. The fact that for this finite interval in $z$ we don't achieve exactly $\phi''(z_{\rm min})=0$ is reflected by the slight non-degeneracy of the minima of the pseudopotential. We illustrate the profiles for solutions near $T_+=T_{+}^{\rm max}$ in Figs.~\ref{Fig:def_prof} and \ref{Fig:def_prof_xi}.

The physical solution with $\phi''(z)\rightarrow0$ at large $|z|$ would correspond to the solutions that were searched for in Ref.~\cite{Konstandin:2010dm}. The solution with a minimal value of  $\phi''(z_{\rm min})$ found here satisfies approximately the constraints derived in the former reference from requiring  a zero driving force (although in fact there is a driving force which is exactly compensated by a nonzero $\vec{F}_{\rm back}$, as illustrated in Fig.~\ref{Fig:static_results}). Defining the parameter 
\begin{align}
 \alpha_c=\frac{l_c}{4 a_+ T_c^4},
\end{align}
where $l_c=T\partial_T(V(h_c,T)-V(0,T)|_{T=T_c})$ is the latent heat of the transition (with $h_c$ the nontrivial VEV at the critical temperature), and with $a_+=\pi^2/30(g_{*,SM}+2N)$ related to the $T^4$ coefficient of the pressure in Eq.~\eqref{eq:pressure}, the following identities from Ref.~\cite{Konstandin:2010dm} are satisfied
\begin{align}\label{eq:KNoconds}\begin{aligned}
v_w^2\sim&\,\frac{1}{6\alpha_c}\,\log\frac{T_{c}}{T_{\rm nuc}},\\
\log\frac{T_c}{T_N}<&\,{\cal O}(1)\left(\sqrt{\frac{\alpha_c}{2
}}-\frac{3}{10}\alpha_c-\frac{1}{5}\alpha_c^{3/2}\right).
\end{aligned}\end{align}

The static profiles for the scalar have a typical width as in Fig.~\ref{fig:PhaseTransition}, $L \sim 20/m_W\sim 30/T$. The local equilibrium approximation is expected to hold if $L/\gamma(v_w)$ is above the mean free path $\lambda_{\rm mfp}$ of particles in the plasma. With $v_w\lesssim0.3$ the Lorentz contraction factor is of order one, while  Ref.~\cite{Liu:1992tn} estimated $\lambda_{\rm mfp}\lesssim \hat m^2_W(T)/(10\pi\alpha_w^2 T^3)$, where $\hat m^2_W(T)$ is the temperature-dependent $W$ mass, and $\alpha_w=g_2^2/(4\pi)$. In our bubbles, we have $h\lesssim1.5 m_W\sim T$, giving $\hat m^2_W(T)\lesssim T^2/9$  and $\lambda_{\rm mfp}\lesssim3/T$. Hence the local equilibrium approximation is indeed justified.

\section{\label{sec:static2} Static planar bubble profiles and consistent detonations}

In the following we apply the same treatment of the previous section to the search of consistent detonation profiles. Given the inverse proportionality between the wall-velocity and the increase of entropy density across the bubble in Eq.~\eqref{eq:vwall_vfront}, one expects higher wall velocities if the phase transition increases the mass of a lower number of particles. This also fits with the proportionality of the backreaction force to the increase in entropy density in Eq.~\eqref{eq:FrictionExact}. As for the choice of couplings described in Section~\ref{sec:model} we found deflagration solutions for $N=4$, we hope to find larger wall velocities (and possible detonation solutions) for $N=2$.

\begin{figure}
   \centering
    \hskip0cm \includegraphics[width=0.38\textwidth]{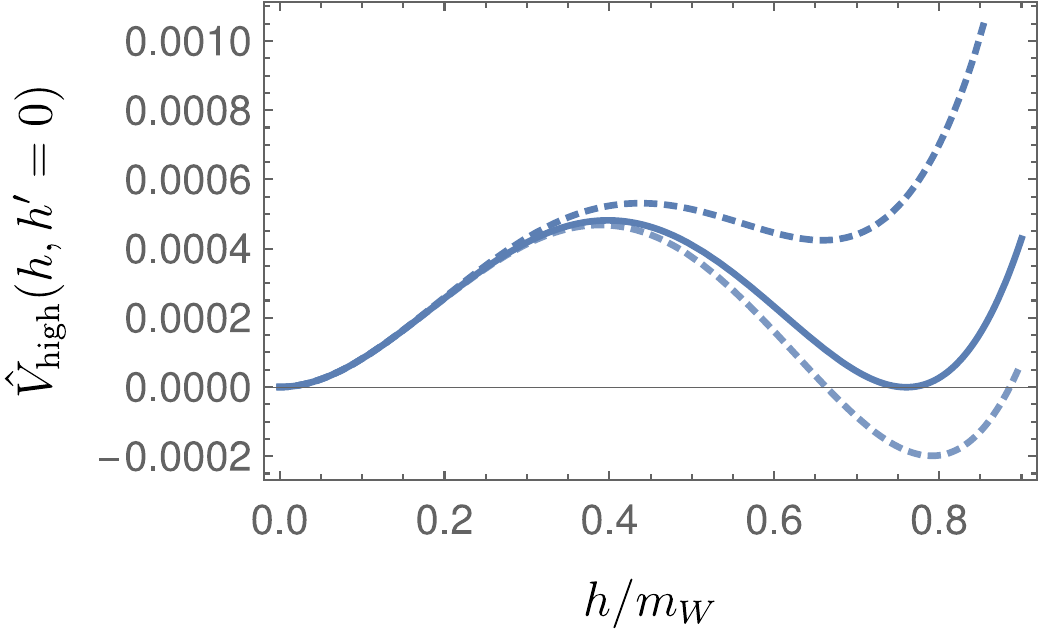}
\caption{Pseudopotential $\hat V_{\rm high}(h,h'=0)$ in the $N=2$ case for $T_+=T_{\rm nuc}=126.229$ GeV, with $v_+$ taking the values (from top to bottom): -0.65, -0.723, -0.80. The central choice of $v_+$ gives a consistent detonation profile.}
\label{fig:Vhathigh}
\end{figure}

\begin{figure}
   \centering
    \hskip0.4cm \includegraphics[width=0.3\textwidth]{fig/hwall_det.pdf}
     \includegraphics[width=0.33\textwidth]{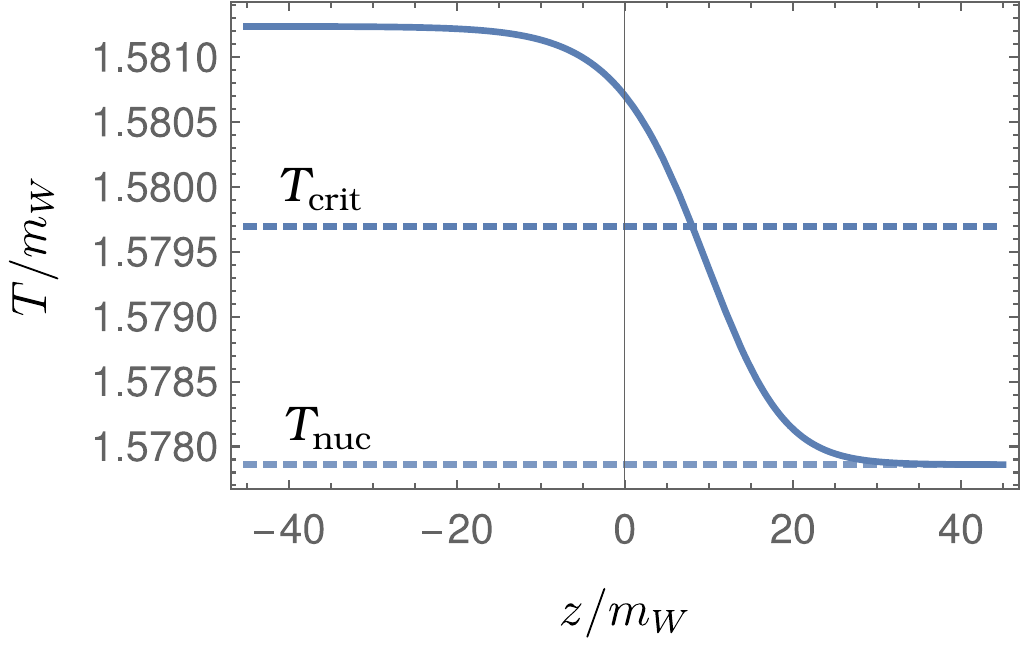}
      \includegraphics[width=0.33\textwidth]{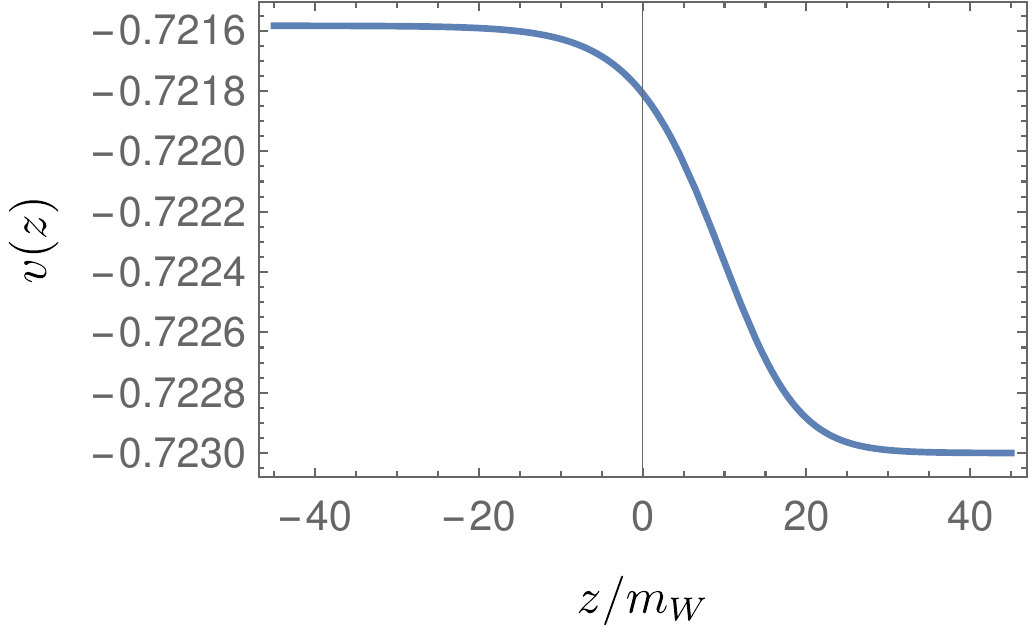}
\caption{Detonation profiles of the Higgs, temperature and velocity across the bubble wall. Note how the fluid velocity increases behind the bubble, and the temperature rises. One reaches $T_->T_c$, but the Higgs is still allowed to be in a metastable minimum. In the hydrodynamic solution far behind the wall, the temperature drops such that the Higgs is stabilized (see Fig.~\ref{Fig:det_prof_xi}).}
\label{Fig:det_prof}
\end{figure}

For this choice we find no acceptable deflagration profile with the techniques of the previous section, despite the fact that the parameters satisfy the condition in the second line of Eq.~\eqref{eq:KNoconds} derived in Ref.~\cite{Konstandin:2010dm} for deflagration profiles in local equilibrium.
On the other hand, by choosing the ``high'' branch of solutions of $T(h,h')$, we find
acceptable detonation profiles. The solutions with $|\phi''(z_{\rm min})|\rightarrow0$ are found for $v_{+}=-0.723$ (and of course $T_+=T_{\rm nuc}=126.229$ GeV for $N=2$), giving a supersonic wall velocity $v_w=0.723$. This can be connected to a detonation profile behind the wall that solves Eq.~\eqref{eq:hydro} with the fluid velocity dropping to zero as expected.  Fig.~\ref{fig:Vhathigh} shows the pseudopotential $\hat{V}_{\rm high}(h,h'=0)$ calculated for three different choices of  $v_+$, the central one giving the physical solution.
The profiles for the Higgs field, the velocity and temperature along the wall are shown in Fig.~\ref{Fig:det_prof}. Note the heating effect behind the bubble, 
with the value of $T$ setting onto $T_-=126.499 \,{\rm GeV}>T_c=126.376$ GeV. For this temperature above the critical one there is still a nontrivial Higgs minimum, yet with a higher energy than the minimum at the origin. The physical interpretation is that the fluid heats immediately behind the bubble, driving the Higgs to a metastable minimum. For solving the hydrodynamic profile behind the bubble, the value of the Higgs can be assumed  not to change much i.e. the energy of the Higgs vacuum shifts with temperature, but the relative changes of the VEV are small. Then assuming a constant Higgs value one can solve the hydrodynamic equations \eqref{eq:hydro} behind the bubble, which confirms that the temperature drops to a value $T_{\rm in}=126.259\,{\rm GeV}<T_c$, so that the Higgs can be stabilized at the absolute minimum of the finite-temperature potential well inside the bubble. The backreaction force computed from Eq.~\eqref{eq:FrictionExact} is found to be $|\vec{F}_{\rm back}|/A/m_W^4=0.648$, while the Eq.~\eqref{eq:Force} gives a result which is 3.2 times larger. 

\begin{figure}
   \centering
     \includegraphics[width=0.33\textwidth]{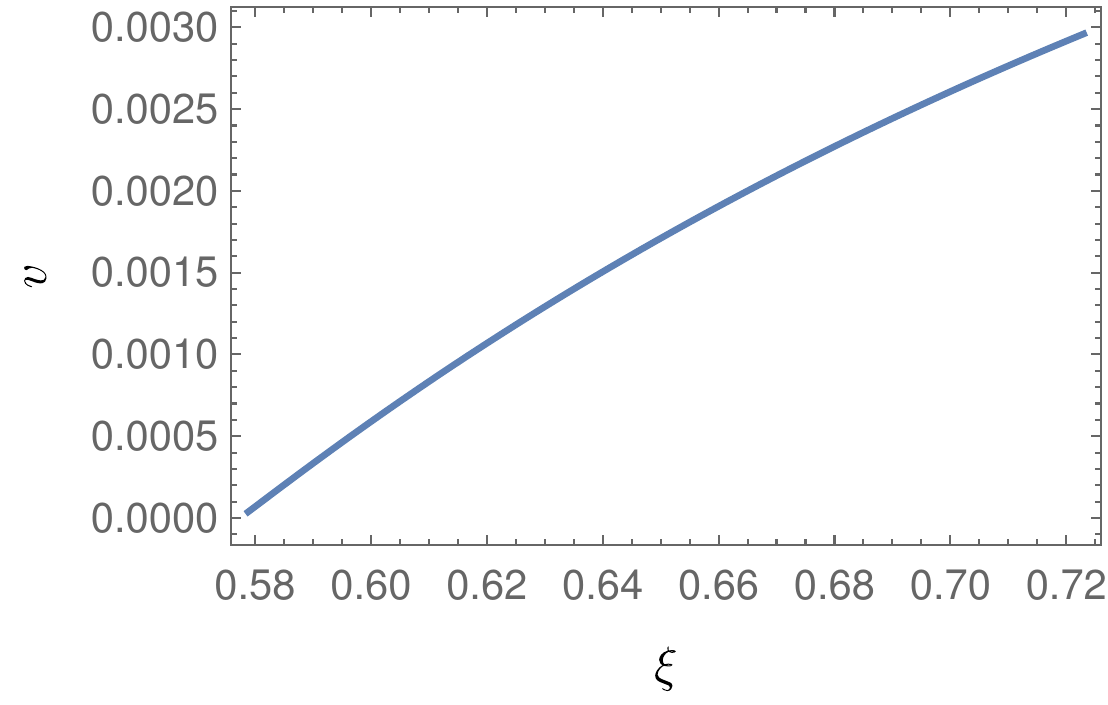}
      \includegraphics[width=0.33\textwidth]{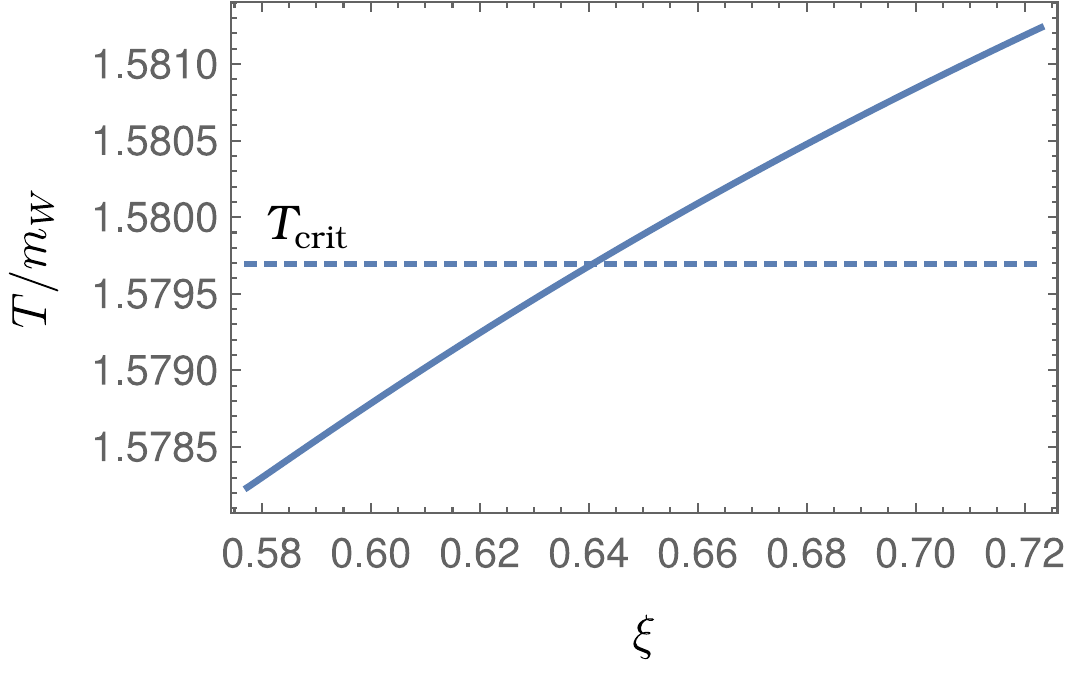}
\caption{Hydrodynamic profiles for the fluid temperature and velocity behind the bubble wall, assuming a constant Higgs value.}
\label{Fig:det_prof_xi}
\end{figure}

\section{\label{sec:conclusions}Discussion and conclusions}

In this work we have confirmed and provided new insights on the hydrodynamic effects that give rise to subluminal bubble propagation in  first-order phase transitions in which equilibrium is maintained locally. Such subluminal propagation in equilibrium has been proposed for deflagrations in Ref.~\cite{Konstandin:2010dm} and for general transitions in \cite{Mancha:2020fzw},  and remains in contrast to the common view that links bubble friction with out-of-equilibrium effects. In our work we have provided an understanding of the subliminal propagation as a consequence of the conservation of the total entropy of the degrees of freedom in local equilibrium: in a simplified planar expansion in which detonation or deflagration fronts develop (which typically propagate subluminally) entropy conservation relates the bubble wall and front velocities. We went beyond the work of Ref.~\cite{Mancha:2020fzw} by clarifying the origin of the friction forces in the differential equations for the scalar field and the temperature and velocity profiles of the plasma, and by calculating the time-dependent bubble expansion  in a SM extension with additional scalars. The slowing down of the bubble arises from terms sensitive to the dependence of the entropy on the scalar field background. These backreaction effects
can be accounted for by using conservation of the stress-energy momentum tensor and incorporating the background-field-dependence of the plasma's pressure and enthalpy, which can be derived straightforwardly from the thermal corrections to the effective potential. 

We have argued that the conservation of the total entropy of the equilibrated degrees of freedom implies that the fluid must heat up in a region near the bubble, which offers a natural connection with the heating effect that was pointed out in Ref.~\cite{Konstandin:2010dm}. That reference analyzed bubble profiles by considering the equations in the static limit, while accounting for consistent hydrodynamic deflagration profiles away from the bubble. We have done analogous computations and found that, while the effect pointed out in Ref.~\cite{Konstandin:2010dm} was assumed to be restricted to deflagrations, one can also get consistent detonation solutions with subluminal wall velocities, as expected from the results of Ref.~\cite{Mancha:2020fzw}.

The computations of the static wall profiles allowed us to estimate the resulting backreaction force, computed from Eq.~\eqref{eq:FrictionExact}, against the results~\eqref{eq:Force}  of Ref.~\cite{Mancha:2020fzw}, which excludes runaway bubbles. We found qualitative agreement up to $\mathcal{O}(1)$ effects related to the change of velocity and temperature across the wall. 

In our calculations we considered a scenario in which the hypothesis of local equilibrium seems to be  justified. Nevertheless, in general settings in which some species remain out of equilibrium, we expect as pointed out in Ref.~\cite{Mancha:2020fzw} that the backreaction force from the equilibrated plasma will still play an important role, as the conservation of the total entropy of the degrees of freedom in equilibrium will typically require subluminal speeds. This effect should be accounted for properly in such cases.

\begin{acknowledgements} The authors thank Jos\'e Miguel No and Thomas Konstandin for discussions. C.T. acknowledges further discussions with Bj\"orn Garbrecht and financial support by DFG through the SFB 1258 and the ORIGINS Cluster of Excellence. M.S. is supported by the STFC under grant ST/P001246/1.

\end{acknowledgements}

\bibliographystyle{mykp}
\bibliography{references}  

\end{document}